\documentclass[dvips]{article}
\usepackage{icrctc07}

\title{Establishing a connection between high-power pulsars and
  very-high-energy gamma-ray sources}
\shorttitle{High-power pulsars and
  very-high-energy gamma-ray sources}
\authors{S. Carrigan$^{1}$, J.A.~Hinton$^{1,2}$, W. Hofmann$^{1}$, K. Kosack$^{1}$, T.
  Lohse$^{3}$ and O. Reimer$^{4}$ for the H.E.S.S. Collaboration}
\shortauthors{S. Carrigan for H.E.S.S.}
\afiliations{ \small
  $^1$Max-Planck-Institut f\"ur Kernphysik, P.O. Box 103980, D 69029 Heidelberg, Germany\\
  $^2$Landessternwarte, Universit\"at Heidelberg, K\"onigstuhl, D 69117 Heidelberg, Germany\\
  $^3$Institut f\"ur Physik, Humboldt-Universit\"at zu Berlin, Newtonstr. 15, D 12489 Berlin, Germany\\
  $^4$Stanford University, HEPL \& KIPAC, Stanford, CA 94305-4085, USA}
\email{svenja.carrigan@mpi-hd.mpg.de}

\abstract{In the very-high-energy (VHE) gamma-ray wave band, pulsar
  wind nebulae (PWNe) represent to date the most populous class of
  Galactic sources. Nevertheless, the details of the energy conversion
  mechanisms in the vicinity of pulsars are not well understood, nor
  is it known which pulsars are able to drive PWNe and emit
  high-energy radiation. In this paper we present a systematic study
  of a connection between pulsars and VHE $\gamma$-ray sources based
  on a deep survey of the inner Galactic plane conducted with the High
  Energy Stereoscopic System (H.E.S.S.). We find clear evidence that
  pulsars with large spin-down energy flux are associated with VHE
  $\gamma$-ray sources. This implies that these pulsars emit on the
  order of 1\% of their spin-down energy as TeV $\gamma$-rays.
}

\begin{document}
\maketitle

In 1989, the Crab Nebula was discovered as the first celestial source
of VHE $\gamma$-radiation \cite{WhippleCrabDisc}. The pulsar inside the nebula drives a
powerful wind of highly relativistic particles that ends in a
termination shock from which high-energy particles with a wide
spectrum of energies emerge \cite{Gaensler06}. High-energy
electrons\footnote{here and in the following, `electrons' refers to both
electrons and positrons} among these particles
can give rise to two components of electromagnetic radiation: a
low-energy component from synchrotron radiation and a high-energy
component from inverse Compton (IC) up-scattering of ambient photons.

Recently, advances in VHE instrumentation have made the discovery of
many new, predominantly Galactic, sources possible. Of these, a
significant number can be identified as PWNe. Prominent examples are
the PWN of the energetic pulsar PSR~B1509$-$58 in the supernova
remnant MSH~15$-$5$2$ \cite{HESS:MSH1552}, and HESS~J0835$-$455 \cite{HESS:velax},
associated with Vela~X, the nebula of the Vela pulsar. 

These $\gamma$-ray PWNe are extended objects with an angular size of a fraction of a
degree, translating into a size of some 10\,pc for typical distances
of a few kpc.  In addition to the open puzzle of pulsar spin-down
power conversion, a surprising observation is that the centroids of
these $\gamma$-ray PWNe are often displaced from their pulsars by
distances similar to the nebular size. Such displacements, although
usually at smaller scales, are also seen in some X-ray PWNe. The
origin of the displacement remains unknown. It might be attributed to
pulsar motion (e.g. \cite{Swaluw04}), causing the pulsar to leave its nebula
behind, or to a density gradient in the ambient medium \cite{blondin01:PWN}.

The aforementioned examples of coincidences between VHE $\gamma$-ray
sources and radio pulsars motivated a systematic search for VHE
counterparts of energetic pulsars using the H.E.S.S. system of imaging
Cherenkov telescopes located in Namibia \cite{Hinton:2004}. To be
detectable by H.E.S.S., a source at distance $d$ has to provide a
$\gamma$-ray luminosity in the 1~TeV to 10~TeV range of $L_\gamma \sim
10^{32}~d^2$~erg\,s$^{-1}$kpc$^{-2}$. Assuming a conversion efficiency
of ~1\% of pulsar spin-down energy loss $\dot{E}$ into TeV
$\gamma$-rays (where $\dot{E}$ is determined from the measurement of
the rotation period $\Omega$ and the rate at which the rotation slows
down $\dot{\Omega}$), PWNe of pulsars with $\dot{E}$ around
$10^{34}~d^2$~erg\,s$^{-1}$kpc$^{-2}$ might be detectable. We note
that for typical electron spectra, only a small fraction of the total
energy in electrons is carried by the multi-TeV electrons, that are
responsible for TeV $\gamma$-rays by IC scattering off ambient photons
(including those from the cosmic microwave background) and for keV
$\gamma$-rays by synchrotron radiation. Even a 1\% energy output in
TeV $\gamma$-rays already implies a large fraction of spin-down energy
loss going into relativistic electrons.

Here we investigate how the probability to detect in VHE $\gamma$-rays
PWNe surrounding known pulsars varies with the spin-down energy loss
of the pulsar, testing the plausible assumption that the $\gamma$-ray
output of a PWN correlates in some fashion with the power of the
pulsar feeding it.

The VHE $\gamma$-ray data set used to search for $\gamma$-ray emission
near the location of known radio pulsars comprises all data used in
the H.E.S.S. Galactic plane survey
\cite{HESS:scanpaper1,HESS:scanpaper2}, including an extension of the
survey to Galactic longitudes $-60^{\circ} < l < -30^{\circ}$,
dedicated observations of Galactic targets and re-observations of
H.E.S.S. survey sources. The search covers a range in Galactic
longitude from $-60^{\circ}$ to $30^{\circ}$ while the range in Galactic
latitude is restricted to $\pm$2\,deg, a region well covered in the
survey. A total of 435 pulsar locations are tested, taken from the
Parkes Multibeam Pulsar Survey (PMPS, \cite{Parkes4} and references
therein), as recorded in the ATNF pulsar catalogue. Pulsars without
measured period derivatives are ignored.  Over the range of the
H.E.S.S. survey, the PMPS provides reasonably uniform sensitivity
\cite{ATNF}, enabling a reliable estimate of the frequency of chance
coincidences between a $\gamma$-ray source and a pulsar. The analysis
of the $\gamma$-ray data follows the standard H.E.S.S. analysis
\cite{HESS:crab}. Initially, a sky map is generated providing the
significance of a $\gamma$-ray excess for a given position. Taking
into account the properties of known $\gamma$-ray PWNe, the search is
optimised for slightly extended sources -- on the scale of the angular
resolution ($\approx$~0.1\,deg) of the H.E.S.S. telescopes -- and
allows for small offsets from the pulsar positions. Each excess is
determined by counting $\gamma$-ray candidate events within $\theta
\leq 0.22$\,deg ($\theta^2 \leq 0.05$\,deg$^2$) of a given position
and subtracting a background estimated from areas in the same field of
view. The sky map is used to look up the significance of a
$\gamma$-ray excess at the position of the radio pulsars, as well as
for randomly generated test positions used to evaluate the statistical
significance of the association (details are given below). We require
an excess significance of at least 5 standard deviations above the
background as a signature of a VHE $\gamma$-ray signal.  Given the
modest number of trials - the 435 pulsar locations - the number of
false detections is negligible with this requirement and in any case
small compared to the probability for chance coincidences between
radio pulsars and VHE $\gamma$-ray sources.

\begin{figure*}[ht]
  \centering
  \resizebox{0.8\hsize}{!}{\includegraphics{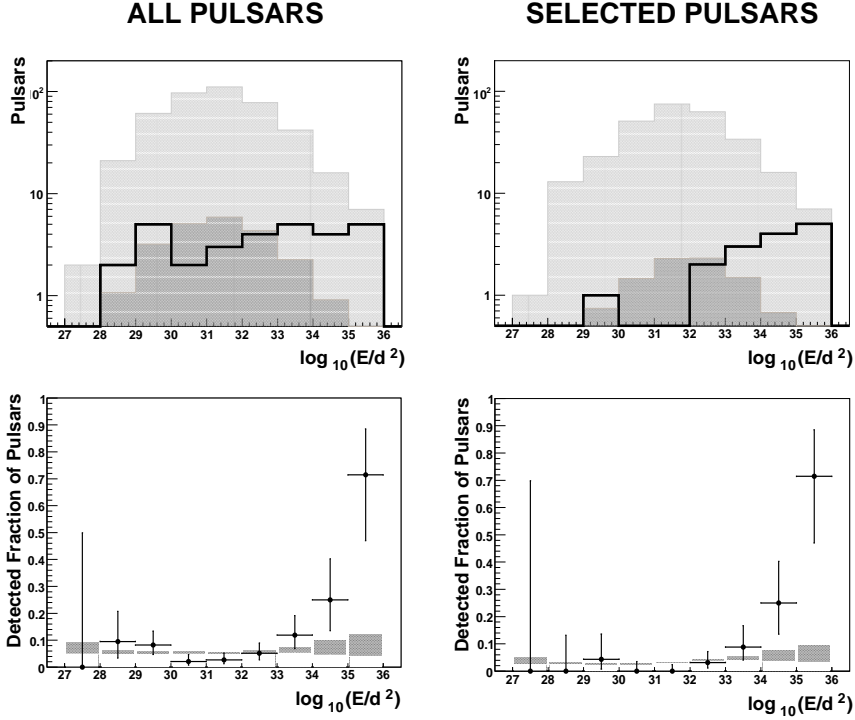}}
\caption{\emph{\bf{Top row:}} Distribution in
log$_{10}(\dot{E}/d^2)$ of all PMPS pulsars in the H.E.S.S. scan range
(shaded in light grey), of chance coincidences (shaded in dark grey)
and of detected pulsars (black line). Here, $\dot{E}/d^2$ is measured
in erg\,s$^{-1}$kpc$^{-2}$. \emph{\bf{Bottom row:}} The points
show the fraction of pulsars with significant $\gamma$-ray excess at
the pulsar position, as a function of log$_{10}(\dot{E}/d^2)$.  
The
shaded band represents the probability for a chance coincidence. The
width of the band accounts for the uncertainty in the width of the
latitude distribution of pulsars.
\emph{\bf{Left:}} all pulsars; \emph{\bf{right:}} double occurrences
of gamma-ray sources removed by omitting pulsars which overlap with
stronger pulsars or known non-pulsar sources.
}
\end{figure*}

Of the 435 pulsars, 30 are found with significant $\gamma$-ray
emission at the pulsar location (Fig.~1, top left panel).  The lower
left panel of Fig.~1 displays the fraction of pulsars with such
$\gamma$-ray emission for different intervals in spin-down flux
$\dot{E}/d^2$. The fraction is about 5\% for pulsars with spin-down
flux below $10^{33}$~erg\,s$^{-1}$kpc$^{-2}$ and increases to about
70\% for pulsars with $\dot{E}/d^2$ above
$10^{35}$~erg\,s$^{-1}$kpc$^{-2}$. Not all of these associations are
necessarily genuine. The rate of chance coincidences is estimated by
generating $10^6$ realisations of random pulsar samples (each
consisting on average of 435 ``pulsars'') following the distribution
in longitude and latitude of the PMPS pulsars and taking into account
the narrowing of the distribution in latitude with increasing
spin-down flux. The expected fraction of chance coincidences is shown
as dark shaded areas in Fig.~1 and varies between 4\% to 12\%. All
associations with pulsars with $\dot{E}/d^2 <
10^{33}$~erg\,s$^{-1}$kpc$^{-2}$ are within statistical errors
consistent with chance coincidences. Indeed for plausible values of
the ratio between the $\gamma$-ray luminosity and the pulsar spin-down
energy loss, $L_\gamma/\dot{E}$, no detectable emission would be
expected from such pulsars.  On the other hand, the detection of
emission from high-power pulsars is statistically significant.  The
probability that the detection of VHE sources coincident with 9 or
more of the total of 23 pulsars above $\dot{E}/d^2 >
10^{34}$~erg\,s$^{-1}$kpc$^{-2}$ results from a statistical
fluctuation is $\sim 3.4 \times 10^{-4}$. For detection of 5 or more
of the total of 7 pulsars above $10^{35}$~erg\,s$^{-1}$kpc$^{-2}$, the
chance probability is $\sim 4.2 \times 10^{-4}$.

Given the high density of pulsars, a single $\gamma$-ray source may
even coincide with more than a single pulsar, and thus appear more
than once amongst the ``detections'' in the upper left panel of
Fig.~1. Removal of such double occurrences (Right panels of Fig.~1)
does not change the conclusion, and none of the high-luminosity
pulsars is affected. Details will be given elsewhere.

The results shown in Fig.~1 demonstrate that a large fraction of
high-luminosity pulsars correlate with sources of VHE $\gamma$-rays,
emitting with a $\gamma$-ray luminosity of order 1\% of the pulsar
spin-down power. The positive correlation does not necessarily imply
that the pulsar or PWN itself is responsible for the $\gamma$-ray
flux. It could also result from some other mechanism correlated with
the pulsar or its creation, such as a supernova shock wave. The
correlation found between $\gamma$-ray detectability and spin-down
flux $\dot{E}/d^2$ argues in favour of a pulsar-related origin of the
$\gamma$-ray signal. On the other hand, for the PMPS pulsar sample,
$\dot{E}/d^2$ also correlates closely with the spin-down age $T$ of
the pulsar, $\dot{E}/d^2 \sim T^{-3/2}$, and obviously with distance
$d$, both parameters relevant for determining the $\gamma$-ray flux
from shock-wave driven supernova remnants.
 
The exact relation between pulsar parameters and $\gamma$-ray
luminosity is an interesting issue. Variations in exposure and hence
in detection threshold over the survey range, as well as the
uncertainty in pulsar distance will smear out the turn-on curve of
detectability versus $\dot{E}/d^2$ shown in Fig.~1, but cannot fully
account for the rather slow turn-on over a range of more than one
order of magnitude in $\dot{E}/d^2$, combined with a detection
probability below unity for even the highest-power pulsars.  This
indicates that $\dot{E}/d^2$ cannot be the only parameter relevant for
the $\gamma$-ray flux.  The same conclusion is obtained from the
observed variation of $L_\gamma/\dot{E}$ of about an order of
magnitude among the detected pulsars. However, this present pulsar
sample is too small to investigate the dependence of $L_\gamma$ on
multiple pulsar parameters, e.g. including pulsar age.

A constant $L_\gamma/\dot{E}$ is also not necessarily expected. For a
given age, the integral energy fed by the pulsar into the PWN
increases with $\dot{E}$.  Apart from expansion losses, pulsar
spin-down power is shared between particle energy and magnetic field
energy. If equipartition between the two energy densities is assumed
\cite{Reynolds84,rees74}, the magnetic field in the PWN will increase
with $\dot{E}$ and hence the energy loss by synchrotron radiation will
increase relative to and at the expense of inverse Compton
$\gamma$-ray production.  Indeed, un-pulsed X-ray luminosity of
pulsars is observed to increase faster then $\dot{E}$, $L_X \propto
\dot{E}^{1.4\pm0.1}$ \cite{cheng04}. In such scenarios, magnetic field
values and therefore the balance between X-ray and $\gamma$-ray
emission will also depend on volume, i.e. on the expansion speed of
the nebula and hence on the ambient medium. In addition, the current
spin-down luminosity $\dot{E}$ may not be the only relevant scale; if
the pulsar age is shorter than or comparable to the electron cooling
time, relic electrons injected in early epochs with higher spin-down
power will still contribute and may enhance $L_\gamma$ significantly
compared to the quasi-steady state achieved for old pulsars.

\small

\section{Acknowledgements}
The support of the Namibian authorities and of the University of Namibia
in facilitating the construction and operation of H.E.S.S. is gratefully
acknowledged, as is the support by the German Ministry for Education and
Research (BMBF), the Max Planck Society, the French Ministry for Research,
the CNRS-IN2P3 and the Astroparticle Interdisciplinary Programme of the
CNRS, the U.K. Science and Technology Facilities Council (STFC),
the IPNP of the Charles University, the Polish Ministry of Science and 
Higher Education, the South African Department of
Science and Technology and National Research Foundation, and by the
University of Namibia. We appreciate the excellent work of the technical
support staff in Berlin, Durham, Hamburg, Heidelberg, Palaiseau, Paris,
Saclay, and in Namibia in the construction and operation of the
equipment.

\bibliography{icrc0493}
\bibliographystyle{plain}

\end{document}